\documentclass[twoside,twocolumn,9pt]{article}
\usepackage{extsizes}
\usepackage[super,sort&compress,comma]{natbib} 
\usepackage[version=3]{mhchem}
\usepackage[left=1.5cm, right=1.5cm, top=1.785cm, bottom=2.0cm]{geometry}
\usepackage{balance}
\usepackage{mathptmx}
\usepackage{sectsty}
\usepackage{graphicx} 
\usepackage{lastpage}
\usepackage[format=plain,justification=justified,singlelinecheck=false,font={stretch=1.125,small,sf},labelfont=bf,labelsep=space]{caption}
\usepackage{float}
\usepackage{fancyhdr}
\usepackage{fnpos}
\usepackage[english]{babel}
\usepackage{array}
\usepackage{droidsans}
\usepackage{charter}
\usepackage[T1]{fontenc}
\usepackage[usenames,dvipsnames]{xcolor}
\usepackage{setspace}
\usepackage[compact]{titlesec}
\usepackage{hyperref}

\usepackage{epstopdf}

\usepackage{pdfpages}

\definecolor{cream}{RGB}{222,217,201}

\begin{document}

\pagestyle{fancy}
\thispagestyle{plain}
\fancypagestyle{plain}{
\renewcommand{\headrulewidth}{0pt}
}

\makeFNbottom
\makeatletter
\renewcommand\LARGE{\@setfontsize\LARGE{15pt}{17}}
\renewcommand\Large{\@setfontsize\Large{12pt}{14}}
\renewcommand\large{\@setfontsize\large{10pt}{12}}
\renewcommand\footnotesize{\@setfontsize\footnotesize{7pt}{10}}
\makeatother

\renewcommand{\thefootnote}{\fnsymbol{footnote}}
\renewcommand\footnoterule{\vspace*{1pt}%
\color{cream}\hrule width 3.5in height 0.4pt \color{black}\vspace*{5pt}} 
\setcounter{secnumdepth}{5}

\makeatletter 
\renewcommand\@biblabel[1]{#1}            
\renewcommand\@makefntext[1]%
{\noindent\makebox[0pt][r]{\@thefnmark\,}#1}
\makeatother 
\renewcommand{\figurename}{\small{Fig.}~}
\sectionfont{\sffamily\Large}
\subsectionfont{\normalsize}
\subsubsectionfont{\bf}
\setstretch{1.125} 
\setlength{\skip\footins}{0.8cm}
\setlength{\footnotesep}{0.25cm}
\setlength{\jot}{10pt}
\titlespacing*{\section}{0pt}{4pt}{4pt}
\titlespacing*{\subsection}{0pt}{15pt}{1pt}

\fancyfoot{}
\fancyfoot[RO]{\footnotesize{\sffamily{\thepage}}}
\fancyfoot[LE]{\footnotesize{\sffamily{\thepage}}}
\fancyhead{}
\renewcommand{\headrulewidth}{0pt} 
\renewcommand{\footrulewidth}{0pt}
\setlength{\arrayrulewidth}{1pt}
\setlength{\columnsep}{6.5mm}
\setlength\bibsep{1pt}

\makeatletter 
\newlength{\figrulesep} 
\setlength{\figrulesep}{0.5\textfloatsep} 

\newcommand{\topfigrule}{\vspace*{-1pt}%
\noindent{\color{cream}\rule[-\figrulesep]{\columnwidth}{1.5pt}} }

\newcommand{\botfigrule}{\vspace*{-2pt}%
\noindent{\color{cream}\rule[\figrulesep]{\columnwidth}{1.5pt}} }

\newcommand{\dblfigrule}{\vspace*{-1pt}%
\noindent{\color{cream}\rule[-\figrulesep]{\textwidth}{1.5pt}} }

\makeatother

\twocolumn[
  \begin{@twocolumnfalse}
\sffamily
\begin{tabular}{m{0 cm} p{17cm} }

 & \noindent\LARGE{\textbf{Surface-acoustic-wave driven silicon microfluidic chips for acoustic tweezing of motile cells and viscoelastic microbeads$^{\ast}$}} \\
\vspace{0.3cm} & \vspace{0.3cm} \\

 & \noindent\large{Shichao Jia,$^\dag$\textit{$^{a,b,c,}$} and Soichiro Tsujino$^{\ddag}$\textit{$^{a,b,}$}} \\

 & \noindent\normalsize{Acoustic tweezers comprising a surface acoustic wave chip and a disposable silicon microfluidic chip are potentially advantageous to stable and cost-effective acoustofluidic experiments while avoiding the cross-contamination by reusing the surface acoustic wave chip and disposing of the microfluidic chip. For such a device, it is important to optimize the chip-to-chip bonding and the size and shape of the microfluidic chip to enhance the available acoustic pressure. In this work, aiming at studying samples with the size of a few tens of microns, we explore the device structure and assembly method of acoustic tweezers. By using a polymer bonding layer and shaping the silicon microfluidic chip via deep reactive ion etching, we were able to attain the acoustic pressure up to 2 MPa with a corresponding acoustic radiation pressure of 0.2 kPa for 50 MHz ultrasound, comparable to reported values at lower ultrasound frequencies. We utilized the fabricated acoustic tweezers for non-contact viscoelastic deformation experiments of soft matter and  trapping of highly motile cells. These results suggests that the feasibility of the hybrid chip approach to attaining the high acoustic force required to conduct acoustomechanical testing of small soft matters and cells. }
\end{tabular}

 \end{@twocolumnfalse} \vspace{0.6cm}
]


\renewcommand*\rmdefault{bch}\normalfont\upshape
\rmfamily
\section*{}
\vspace{-1cm}


\footnotetext{\textit{$^{a}$~PSI Center for Life Sciences, Paul Scherrer Institut, Forschungsstrasse 111, Villigen-PSI, 5232, Switzerland.}}
\footnotetext{\textit{$^{b}$~Swiss Nanoscience Institute, University of Basel, Klingelbergstrasse 82, Basel, 4056, Switzerland.}}
\footnotetext{\textit{$^{c}$~Biozentrum, University of Basel, 4056, Switzerland.}}
\footnotetext{\textit{$\dag$~Electronic mail: shichao.jia@psi.ch}}
\footnotetext{\textit{$\ddag$~Electronic mail: soichiro.tsujino@psi.ch}}

\footnotetext{$\ast$~Electronic Supplementary Information (ESI) available. See DOI: 00.0000/00000000.}



\section*{Introduction}

Acoustic tweezers (AT) utilizing acoustic radiation pressure have been intensely studied recently for the contactless handling of various types and sizes of samples in microfluidic (MF) channels, such as polymer particles,\cite{Destgeer:2013, Schwarz:2015, Collins:2016}, droplets\cite{Leibacher:2015, Collins:2013}, cells\cite{Ding:2014, Drinkwater:2020, Leibacher:2015} and more; AT are also more versatile, compared to optical tweezers \cite{Guck:2001, Cojoc:2004, Lee:2023, Dholakia:2020} or magnetic methods,\cite{Wang:1993, Bausch:1999} because samples do not have to be optically transparent or magnetic.\cite{Ozcelik:2018, Bruus:2014, Rufo:2022, Rasouli:2023} The experimental purposes include but are not limited to sorting particles and cells based on their size and/or mechanical properties,\cite{Franke:2010, Jakobsson:2014, Li:2015, Wu:2019, Gao:2021, Wu:2021} trapping,\cite{Li:2020, Baudoin:2019, Baudoin:2020, Doinikov:2020} and patterning populations of cells \cite{Li:2014, Guo:2015, Collins:2015, Deshmukh:2022} or particles.\cite{Shi:2009, Deng:2023} AT have been also successfully adopted for studying motile cellular samples, \cite{Cui:2021, Cui:2023} motility-based sorting of human sperm cells,\cite{Misko:2023} and acoustic stimulation of cells such as \textit{C. elegans}.\cite{Zhou:2017} Conversely, motile cells were also used to characterize an acoustofluidic device\cite{Kim:2021}. 

Matching the acoustic wavelength to the sample size helps achieve a strong acoustic radiation force. For samples with the size of a few tens of microns (a typical size of cells), the required ultrasound frequency is several tens of MHz in aqueous fluid, and such ultrasound is most conveniently generated by surface acoustic wave (SAW) devices which are essentially interdigitated transducers (IDT) fabricated on a piezoelectric substrate.\cite{Gao:2021, Rasouli:2023} To combine the IDT with MF channels housing samples for acoustofluidic experiments, several approaches have been reported. MF channels can be fabricated directly in the same SAW chip, \cite{Zhang:2020} or they can also be prepared by bonding polydimethylsiloxane (PDMS) \cite{Ding:2014, Wu:2021} frames to the SAW chip. 

A similar but alternative approach is to fabricate the MF channel in a separate chip that is to be mounted or bonded to the SAW chip.\cite{Guo:2015, Witte:2014, Hodgson:2009} In the hybrid approach utilizing a glass or silicon MF chip, high acoustic pressure of a few MPa has been reported at ultrasound frequency of $\sim$20 MHz and below for transferring droplets \cite{Hodgson:2009}, focusing particles, \cite{Witte:2014} and microcentrifugation,\cite{Wong:2019} wherein water was used as the couplant between the SAW chip and the MF chip. Since SAW chips are commonly fabricated from a costly piezoelectric substrate such as lithium niobate, the hybrid approach makes the costly SAW chip reusable. At the same time, the low-cost glass or silicon MF chips are  disposable, avoiding the cross-contamination between experiments.\cite{Hodgson:2009, Qian:2021} Using rigid materials such as glass or silicon wafers \cite{Hodgson:2009, Qian:2021} to fabricate MF channels is also amicable for high Q-factor acoustic resonance inside the channel \cite{Laurell:2007} and will potentially compensate for the loss in acoustic coupling between the two chips. In contrast, when MF channels are formed in a PDMS frame bonded to the SAW chip, the reusability of SAW chips appears to be low.\cite{Guo:2015} Also, with PDMS, not only is high-Q acoustic resonance difficult to achieve but also the acoustic loss can increase the temperature of the fluid.\cite{Cui:2021, Cui:2023}  

\begin{figure*}[h]
\centering
  \includegraphics[width=\linewidth]{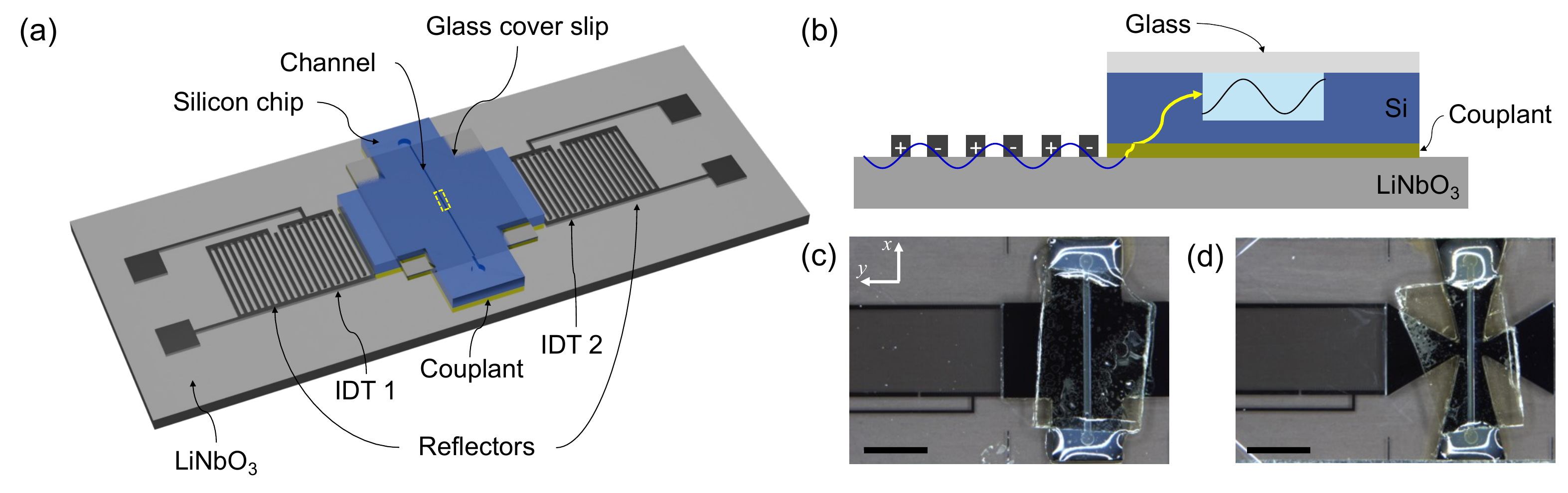}
  \caption{Schematic oblique view (a) and cross-sectional view (b) of the fabricated acoustic tweezers (AT) comprising a LiNbO$_3$ SAW chip and a disposable Si microfluidic (MF) chip. The field of view in Fig.~\ref{fig:2} is marked by a rectangle (yellow dashed line) in (a). (c) and (d) show the top view of the assembled type-1 AT and type-2 AT, respectively. On top of the MF chip, a glass plate caps the channel, of which the two ends were sealed by epoxy glue after infusing the fluid containing samples. SAW is generated by the IDTs (only one side is shown) and excites the MF channel of the Si chip. For the type-2 AT, both sides of the Si chip were shaped into acoustic horns. Scale bars in (c) and (d) are 2 mm.}
  \label{fig:1}
\end{figure*}

Fluid couplants, such as water \cite{Hodgson:2009, Witte:2014, Wong:2019} and silicone grease\cite{Melde:2023} have been shown to allow for quick assembly and disassembly, easing the reuse of the SAW chip and ideal for rapid testing. However, they are plagued by problems such as the chip-to-chip stability (especially at high acoustic power\cite{Qian:2020}) as well as the evaporation and degradation over prolonged experiments.\cite{Qian:2020, Witte:2014, Langelier:2012} These can be mitigated by using a solid couplant such as ultraviolet (UV)-curable epoxy resin.\cite{Langelier:2012} 
In the meantime, the separate fabrication of the MF chip potentially permits further enhancement of acoustic coupling by the exploration of the size and shape of the MF chip, which can be facilitated via deep reactive ion etching (DRIE) in the case of silicon (Si) MF chips but remains rarely investigated to date.

From such a background, we develop hybrid AT comprising separately fabricated SAW chips and Si MF chips. In particular, We explore the optimization of the acoustic coupling between two chips in terms of the chip-to-chip bonding method and the silicon MF chip shape in order to achieve a high acoustic pressure inside the MF channel. The developed AT were applied for trapping highly motile cells and viscoelastic deformation experiments of soft matter, demonstrating the feasibility of the proposed hybrid AT approach.

\section*{Materials and methods}

\subsection*{Device fabrication}

Figure~\ref{fig:1} shows the schematic of the hybrid AT consisting of a SAW chip fabricated from a LiNbO$_3$ wafer and a MF chip fabricated from a Si wafer, together with the microscopic images of the fabricated type-1 (Fig.1(c)) and type-2 (Fig.2(d)) AT.  The ultrasound generated by the IDT in the form of Rayleigh surface acoustic wave is injected into the MF channel via the couplant and the silicon body of the MF chip. 

To fabricate SAW chips, we patterned the interdigitated electrodes on a 4-inch LiNbO$_3$ wafer using photolithography and wet-etching. The LiNbO$_3$ wafer (127.86$^{\circ}$ $Y$-cut and $X$-propagating) is 350 $\mathrm{\mu}$m thick and single-side mirror-polished. First we evaporated a 200 $\mu$m thick Al (with 1\% Si) film on the wafer. Then we spin-coated the wafer with the photoresist S1813 G2 (Microposit, Dow Inc.). Next we loaded the wafer into a direct laser writer (DWL66+, Heidelberg Instruments) and exposed a pattern of the etching mask for the electrodes at a point resolution of 1 $\mathrm{\mu}$m. The unloaded wafer was developed, and immersed in an Al etchant to pattern the Al film. Finally, we diced the wafer into chips with a size of 4 mm $\times$ 7 mm, The interdigitated electrodes consist of 25 or 35 pairs of 20 $\mathrm{\mu}$m wide fingers, each 2.5 mm long, with a 20 $\mathrm{\mu}$m separation between adjacent fingers. On each chip, there are two facing interdigitated electrodes separated by 5.02 mm. They are flanked by reflectors with 50 or 70 fingers, each 20 $\mathrm{\mu}$m wide and spaced 20 $\mathrm{\mu}$m apart.  

We fabricated MF chips from a 4-inch Si and 250 $\mathrm{\mu}$m thick (100) wafer using photolithography and DRIE. The masks for the photolithography were fabricated on chromium coated glass plates by photolithography using the direct laser writer. We used a mask aligner (BA/MA-6, Karl S{\"{u}}ss) to define the etching mask for DRIE. Firstly, we patterned the MF channels using DRIE with SPTS Rapier Deep Reactive Ion Etcher (SPTS Technologies Ltd., United Kingdom) with the photoresist SPR220-3.0 (Megaposit, Dow Inc.). The MF channels have a width of either 200 $\mathrm{\mu}$m or 500 $\mathrm{\mu}$m, a depth of 120 $\mathrm{\mu}$m, and a length of 5 mm. 
Next, we etched through the silicon wafer by DRIE so as to define individual chips. As shown in Fig. 1(c), the side of the type-1 MF chip matches the aperture length (2.5 mm) of the IDT. For the type-2 MF chips, both sides were shaped as acoustic horns as shown in Fig.~\ref{fig:1}(d). The height of the chip on the side in the $x$-direction was reduced from 2.5 mm to 200 $\mathrm{\mu}$m. The narrowest part of the horn is 100 $\mathrm{\mu}$m away from the edge of the MF channel. The processed wafer was finally diced into individual MF chips.  

\subsection*{Assembly of the acoustic tweezers}

The assembly of the AT involves the bonding of the MF chip to the SAW chip and the capping and sealing of the MF channel. We have tested several methods to improve the acoustic pressure in the MF channel, the stability of the AT experiments, and the optical quality. In the first approach, we attached the MF chip on the SAW chip by silicone grease, and capped the MF channel by a piece of transparent Scotch tape (3M Company). 

Since this chip mounting method caused instability at high voltage SAW excitation,  we switched to a polymer (photoresist) for bonding. In this case, after placing a MF chip on a SAW chip, we dispensed a photoresist (AZ nLOF 2020) between the chips via capillary penetration. \cite{Langelier:2012} This was followed by baking on a hot-plate at 110$^{\circ}$C for 2 min. Next, we cross-linked the photoresist by UV exposure (90 s of 365 nm light at 10 mW cm$^{-2}$). The final thickness of the photoresist was $\sim$0.5 $\mathrm{\mu}$m.

To improve the optical resolution, we replaced the tape with a cleaved glass coverslip for capping the MF channel. Firstly, a 5-mm-square, 130-170 $\mathrm{\mu}$m thick coverslip was spin-coated with a photoresist (AZ nLOF 2020), soft-baked at 75 $^{\circ}$C for 30 s, and cleaved into 5-mm-square pieces. Next, a cleaved piece was bonded to the MF chip by heating on a hot-plate at 85 $^{\circ}$C for 2 min while pressing the glass piece with tweezers. Then, we crosslinked the photoresist by UV  exposure (1 min), which was followed by baking the assembly at 110 $^{\circ}$C (2 min) and flushing remover (AZ MIF 726) through the MF channel to remove the photoresist underneath the glass piece within the MF channel. The UV exposure was done on a mask aligner while covering the MF channel. We finally mounted the AT on a custom-built AT holder, wire-bonded IDTs, and infused samples into the MF channel. To avoid evaporation of water, we sealed the inlets of the MF channel with an epoxy glue (Araldite, Huntsman Corp.).     

\subsection*{Preparation of tracers and samples}

To characterize the distribution of the acoustic radiation pressure in the MF channel, we used green fluorescent polyethylene (PE) particles (Cospheric LLC) with a diameter of 3.0 $\pm$ 0.8 $\mathrm{\mu}$m as tracers. The density and the speed of sound of these particles are nominally 1.30 g/cm$^3$ and 2.5${\times}10^3$ m/s, \cite{Ortega-Sandoval:2024} respectively.  We dispensed the particles in de-ionized (DI) water containing 0.01\% Tween 20 with a particle concentration of 0.8 mg/ml and infused the suspension into the MF channel. 

We used \textit{Tetrahymena} (IZB Einzeller Shop, Institut f{\"{u}}r Zellbiologie, Bern, Switzerland) as a motile cell sample, which is a unicellular eukaryote known for their high motility. We infused a suspension of cells into the MF channel and sealed the channel in the same way as described above.

We tested agarose hydrogel beads of non-crosslinked 4\% agarose with a nominal diameter of 50--70 $\mathrm{\mu}$m (Abbexa, LTD) as an example of viscoelastic material. The suspension of the beads was prepared in a similar way as the PE particles.

\subsection*{Setup of acoustic tweezing experiment}

The acoustic tweezing experiment used a custom-built microscope: long-focal-distance objective lenses (Mitsutoyo) with magnifications of 10$\times$ or 20$\times$, a CCD Camera (Grasshopper 3 GS3-U3-28S4M, Teledyne FLIR LLC), and a UV LED (365 nm at $\sim$1 W/mm$^2$, Inolux Corp.) or a white LED for illumination. The UV illumination was done in-line using a dichroic mirror to couple the light to the sample and only the visible light (longer than 450 nm) was recorded for fluorescence imaging. For white LED illumination, the dichroic mirror was replaced with a half mirror. 

To generate ultrasound pulses, RF pulses were applied to IDTs by combining function generators (SG384, Stanford Research Systems), RF amplifiers (VBA1000-18, Vectawave Technology Limited, and ZHL-20W-13X+, Mini-Circuits), and a delay generator (QC9516, Quantum Composer). The first order SAW generated by the IDTs is at $\sim$50 MHz as designed. This was in agreement with the transmission measurement for the assembled AT device using one IDT as the transmitter and the other as the receiver. The width of the output frequency spectrum was found to be $\sim$10 MHz (full width at half maximum), in agreement with the number of the electrodes.\cite{Datta:1986} To identify an acoustic resonance in the MF channel, we therefore scanned the driving frequency around 50 MHz. In the following experiment, we set the driving frequency at 49.2 MHz, which corresponds to one of the strongest resonances. The detailed frequency characteristics will be described elsewhere. 

\subsection*{Particle analysis}

Particle trajectories in the recorded microscopic images were tracked using the plugin TrackMate \cite{Tinevez:2017,Ershov:2022} in the open-source image processing package Fiji.\cite{Schindelin:2012} We used the same method to analyze the motion of \textit{Tetrahymena}. 

\subsection*{Evaluation of acoustic pressure and resonance in MF channels}

To evaluate the acoustic field in the MF channel, we analyzed the motion of the PE tracer particles. During the ultrasound pulse (> 10 ms), the motion of the tracer is in steady state with the acoustic force $F_{\mathrm{ac},y}$ on each tracer being locally balanced with the Stokes drag force $F_{\mathrm{ac},y}$,\cite{Bruus:2014} 
\begin{equation}
    F_{d,y} = -6\pi\eta{r}v_y .
\label{eq:stokes}
\end{equation}
where $\eta$ is the dynamic viscosity of the fluid (equal to  1.00 mPa$\cdot$s in water), $r$ = $\sim$1.5 $\mu$m is the radius of the tracer, and $v_y$ is the velocity of the tracer as evaluated from the displacement between two consecutive frames multiplied by the frame rate (25 fps). 

Since the tracers were much smaller than one half of the ultrasound wavelength, $\lambda/2$ = 15 $\mu$m, $F_{\mathrm{ac},y}$ can be expressed by\cite{Bruus:2014}
\begin{equation}
    F_{\mathrm{ac},y} = 4\pi{\alpha}kr^3E_{\mathrm{ac}}{\mathrm{sin}}(2ky),
\label{eq:Frad}
\end{equation}
where $\alpha$ is the acoustic contrast factor, $k$=2$\pi$/$\lambda$ is the wavenumber of the corresponding acoustic mode in the $y$ direction, and $E_{\mathrm{ac}}$ is the acoustic energy density. $\alpha$ is equal to 0.327 for PE tracers in water according to the following equation,\cite{Bruus:2014} 
\begin{equation}
    \alpha(\Tilde{\rho}, \Tilde{\kappa}) = \frac{1}{3}\left(\frac{5\Tilde{\rho} - 2}{2\Tilde{\rho} + 1} - \Tilde{\kappa}\right),
\label{eq:phi}
\end{equation}
with $\Tilde{\rho}=\rho_{\mathrm{p}}/\rho_0$, and $\Tilde{\kappa}=\rho_{0}c_{0}^2/\rho_{\mathrm{p}}c_{\mathrm{p}}^2$. \cite{Augustsson:2010} $\rho_0 = 1000$ kg/m$^3$ and $\rho_p = 1300$ kg/m$^3$ are respectively the densities of water and PE. $c_0 = 1483$ m/s and $c_p = 2500$ m/s are respectively the speed of sound in water and PE. \cite{Ortega-Sandoval:2024} $E_{\mathrm{ac}}$ in water is given by $p_{ac}$ as
\begin{equation}
    E_{\mathrm{ac}} = \frac{p_{\mathrm{ac}}^2}{4\rho_0{c_0^2}}.
\label{eq:Eac}
\end{equation}
We evaluated $p_{\mathrm{ac}}$ from the measured force distribution and Eqs.~\eqref{eq:Frad}--\eqref{eq:Eac}.

\section*{Results and discussion}

\subsection*{Enhancement of the acoustic radiation pressure by acoustic horn}

First, we examine the distribution of the acoustic radiation pressure in the MF channels of the fabricated ATs shown in Fig.~\ref{fig:1} (c)-(d). The MF channels in these devices are 200 $\mathrm{\mu}$m wide and 120 $\mathrm{\mu}$m deep. As described in Materials and Methods, we set the driving frequency to 49.2 MHz, applied three RF pulses to the right-side IDT of the type-1 AT device (Fig.~\ref{fig:1} (c)), and recorded the motion of the tracer particles. Fig.~\ref{fig:2} (a) (also shown in Supplementary Video 1) displays the aggregation of particles along the pressure nodal lines of the standing wave in the MF channel after three RF pulses with an amplitude of 0.75 V$_{\mathrm{rms}}$, a pulse duration of 100 ms, and a period equal of 1 s. The result is shown in Fig.~\ref{fig:2}(c) by red circular markers for the type-1 AT device. 

\begin{figure}[h]
\centering
  \includegraphics[width=\linewidth]{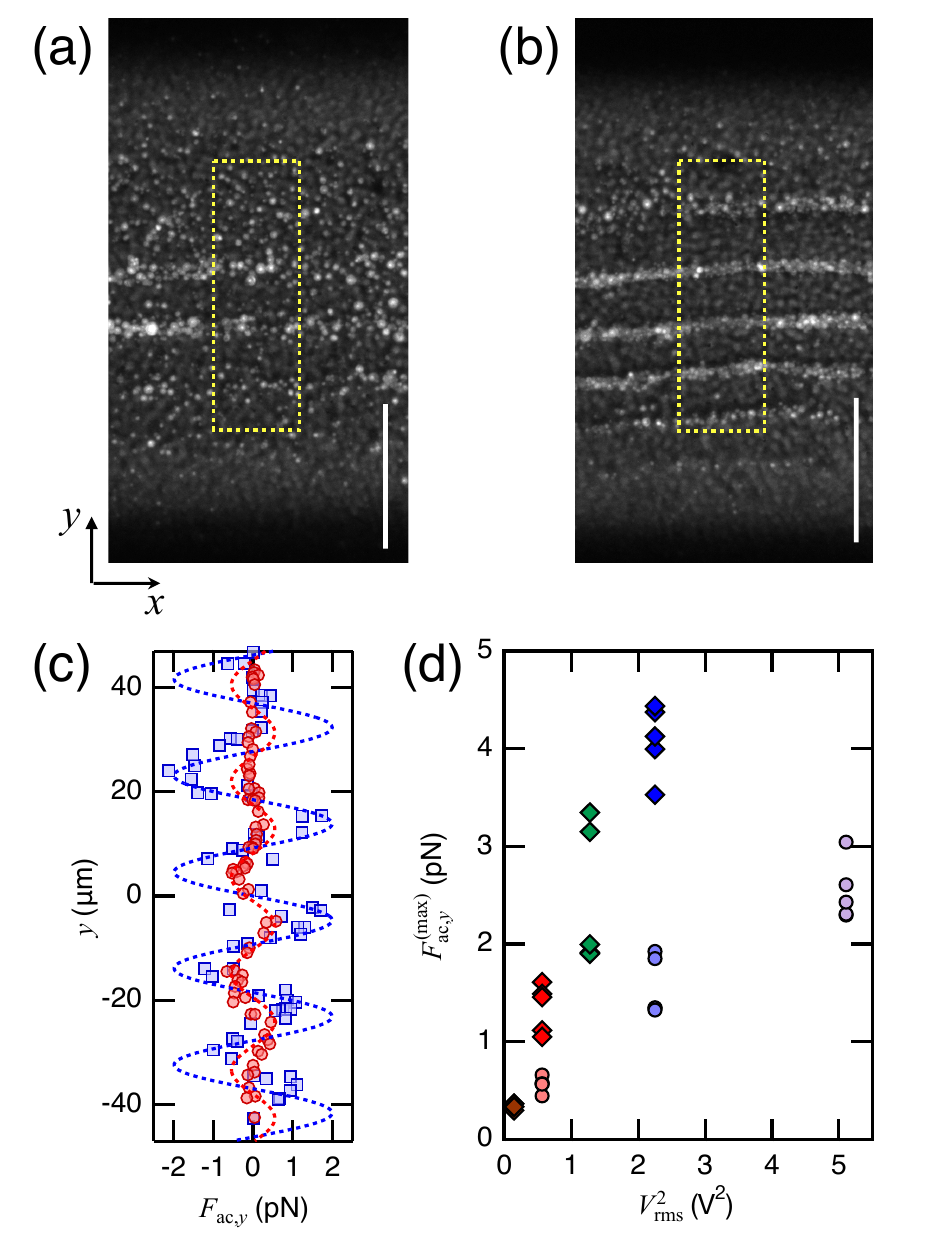}
  \caption{Tracer particle distributions observed in type-1 AT device (a) and type-2 AT device with horn (b), after 49.2 MHz RF pulses (three 100 ms pulses with a period of 1 s and an amplitude of 0.75 $V_{\mathrm{rms}}$) were applied to the left-side IDTs. The higher acoustic radiation pressure in the MF channel in the type-2 AF device is apparent given the clearer aggregation of the tracer particles in (b). Scale bars in (a) and (b) are 50 $\mathrm{\mu}$m. (c) Summary of the evaluated force distribution with an RF voltage of 0.75 $V_{\mathrm{rms}}$. These are evaluated in the ROIs marked by the yellow rectangular frames in (a) and (b). The sinusoidal curves are the fits for the estimated force distribution extracted from the results in the type-1 and type-2 AT devices. (d) Estimated maximum acoustic radiation force as a function of the driving voltage to the IDT for the type-1 AT device (filled circles, without horn) and type-2 AT device (filled diamonds, with horn).}
  \label{fig:2}
\end{figure}

The experiment was repeated with the same RF pulses on the type-2 AT device with the acoustic horns. The results are presented in Fig.~\ref{fig:2}(b) and (c) by blue square markers. The narrower and more concentrated aggregation of tracer particles is evident (see Supplementary Video 2), indicating the enhanced acoustic radiation pressure in the type-2 AT device. This enhancement is attributed to the partial concentration of the incident acoustic radiation within a limited region of the MF channel (see Supplementary Figure S1). Fig. 2(d) illustrates the relationship between the maximum acoustic force and the square of the driving voltage for both types of ATs, revealing an approximately linear increase of the force with increasing $V_{\mathrm{rms}}^2$. The difference in slope and the comparison of the sinusoidal fit amplitudes in Fig. 2(c) show that the acoustic horn is effective to increase the acoustic radiation pressure in the MF channel by a factor of $\sim$5. 

We found that the enhancement of the acoustic radiation pressure is within $\sim$200 $\mu$m along the MF channel, as shown in Supplementary Figure S1. This is expected from the dimension of the end of the horn. 

\subsection*{Influence of the different bonding methods on the acoustic field in the MF channels}

\begin{figure}[h]
\centering
  \includegraphics[width=\linewidth]{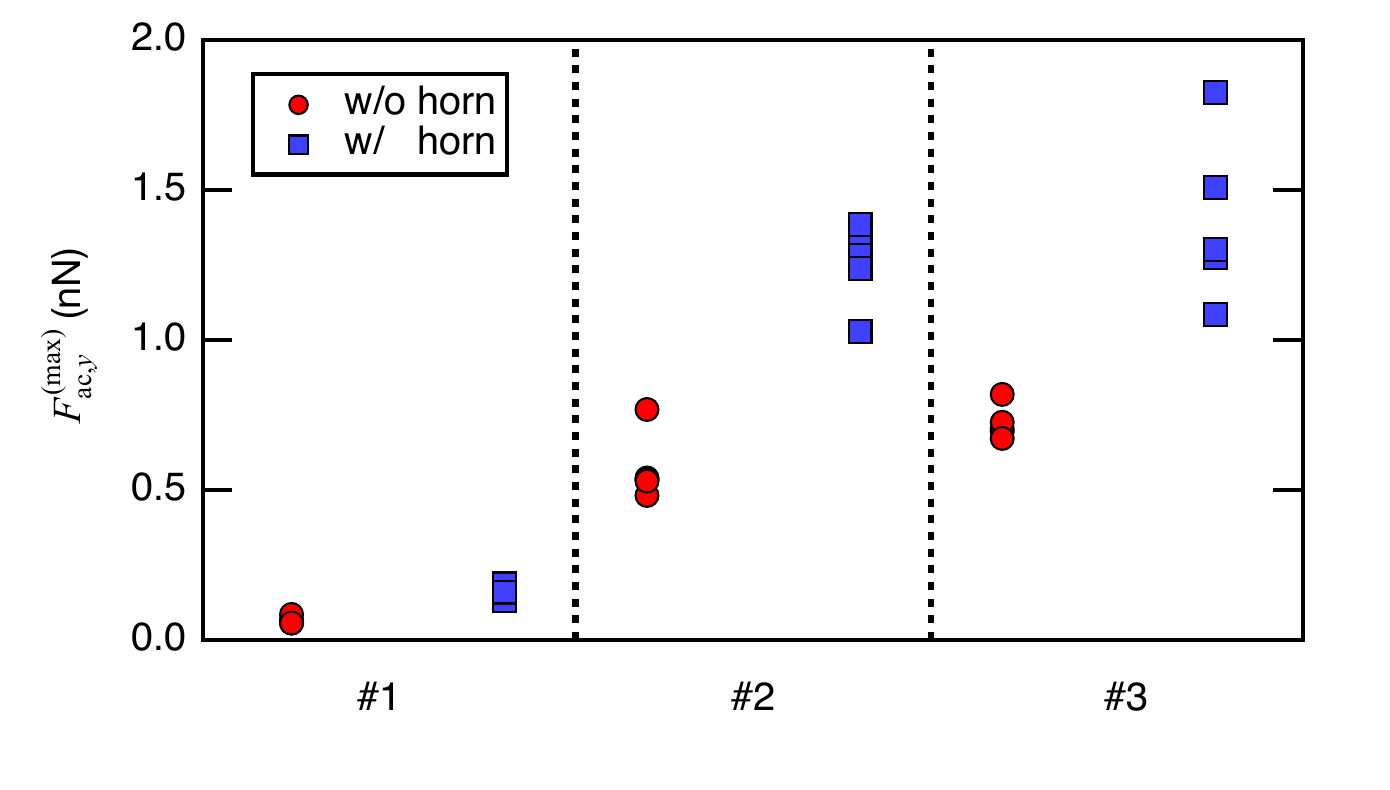}
  \caption{Increase in the maximum acoustic radiation force on 3 µm particles over three generations \#1--\#3 with different technical improvements. Generation \#1: 25-pair electrodes, 50-line reflectors, 41.0 $V_{\mathrm{rms}}$ driving voltage, couplant = silicone gel, Si MF chip: NOT etched through; Generation \#2: 35-pair electrodes, 70-line reflectors, 37.5 $V_{\mathrm{rms}}$ driving voltage, couplant = silicone gel; Si MF chip: NOT etched through; Generation \#3: 35-pair electrodes, 70-line reflectors, 37.5 $V_{\mathrm{rms}}$ driving voltage, couplant = AZ nLOF 2020; Si MF chip: etched through.}
  \label{fig:3}
\end{figure}

To compare the influence of the different chip bonding methods, the SAW chip structure, and the Si-chip shape, the acoustic force observed in three generations of AT, denoted as \#1, \#2 and \#3, are summarized in Fig.~\ref{fig:3}. Here, we display the maximum forces exerted on PE tracer particles, which were observed when the tracer particles were near the pressure peak lines (see Fig. 2) and the SAW chips were driven at 37-41 $V_{\mathrm{rms}}$. Both types of MF chips (with and without acoustic horns) are included in each generation. It is worth noting that, while the Si chips of \#3 were etched through as shown in Fig. 1, the substrates of the Si chips of \#1 and \#2 were etched only partially on the backside with a depth of $\sim$180 $\mathrm{\mu}$m.

The five time increase of the maximum acoustic force from \#1 to \#2 was the result of increasing the number of the IDTs and reflector strips by 30\%, and the improved chip-to-chip alignment by adding alignment markers on the SAW chip. For \#1 and \#2, we used silicone gel (High vacuum grease, DOW Corning Corporation, USA) as the couplant. 

Further increase of the maximum acoustic force by 10--30\% from \#2 to \#3 was the result of 1) replacing the silicone gel with a UV-crosslinked polymer (photoresist) as the couplant, and 2) etching through the Si chips as described in Materials and Methods. We found that the solid couplant and the bonding method as described in Materials and Methods are highly beneficial for the reproducibility of results and long-term stability of devices. We note that through \#1--\#3, the acoustic horn structure is effective to enhance the maximum force by 2--3 times in comparison with the straight Si MF chip on type-1 AT devices. 

\subsection*{Acoustic pressure and resonance mode in MF channels}

To evaluate the acoustic pressure and acoustic modes in the MF channels, we first analyzed the tracer particle trajectories and calculated the force distribution.
The force distribution in Fig.~\ref{fig:2} (c) of the type-2 AT device is fitted to Eq.~\eqref{eq:Frad}. From the amplitude of the fitting curve, we found that, when the RF voltage amplitude $V_{\mathrm{RF}}$ was equal to 0.75 $V_{\mathrm{rms}}$, $E_{\mathrm{ac}}$ = 0.84 J/m$^3$ and $p_{\mathrm{ac}}$ = 86 kPa. In contrast, $p_{\mathrm{ac}}$ = 45 kPa on the type-1 AT device is about one half the value on the device with the acoustic horn. 
By analyzing the results in Fig.~\ref{fig:3} in the same way, we found that $p_{\mathrm{ac}}$ was equal to 2.0 $\pm$ 0.3 MPa on the type-2 AT device with the acoustic horn in \#3 with the maximum force of 1.4 $\pm$ 0.4 nN (when $V_{\mathrm{rms}}$ was 37 V). These analyses indicate that the use of the acoustic horn, together with the optimized chip-to-chip bonding method, is effective to efficiently inject high acoustic energy into the MF chip. 

The period of the sinusoidal fitting curves in Fig.~\ref{fig:2} and Eq.~\eqref{eq:Frad} indicates that the acoustic period, $2\pi/k$, is equal to 37 $\mathrm{\mu}$m and longer than 30 $\mathrm{\mu}$m, namely, the wavelength of 49.2 MHz acoustic wave in water. This suggests that the excited acoustic mode forms a standing wave not only in the horizontal direction but also in the depth direction of the MF channel. Indeed, we found that, by adjusting the focus of the microscope, the tracer particles are aggregated along the horizontal nodal line, approximately 40 $\mathrm{\mu}$m above the bottom of the 120 $\mathrm{\mu}$m deep MF channel. This shows that a standing wave is also formed in the depth direction with a mode number larger than 1. 

Further optimization of the focusing structure appears to be feasible, e.g., by increasing the aspect ratio $r_{h}$ (the ratio of the horizontal length to the vertical length) from $\sim$1 in our acoustic horn to larger values up to 5-10 \cite{Buehling:2022, Inui:2021} even though a focusing structure with $r_h$~$\sim$~1 was also reported.\cite{Meacham:2004}. However, the incident acoustic pressure in our case is a surface wave along the bottom of the Si chip and distributes over the length of the horn. In this configuration, the pressure enhancement is likely to be much smaller than the ratio of the incident area to the area of the exit plane when the input is only through the edge surface.\cite{Kinsler:2000} Also, from two-dimensional cross-sectional finite element simulation, we conclude that, in addition to the acoustic resonance in the MF channel, the acoustic resonance within the silicon could be even more important. Therefore, more detailed understanding of the acoustic mode on the assembled AT device, together with the optimization of the horn shapes, will be essential to further improve the acoustic transmission from the SAW chip into the MF channel.

\subsection*{Synchronous excitation of two IDTs}

\begin{figure}[h]
\centering
  \includegraphics[width=\linewidth]{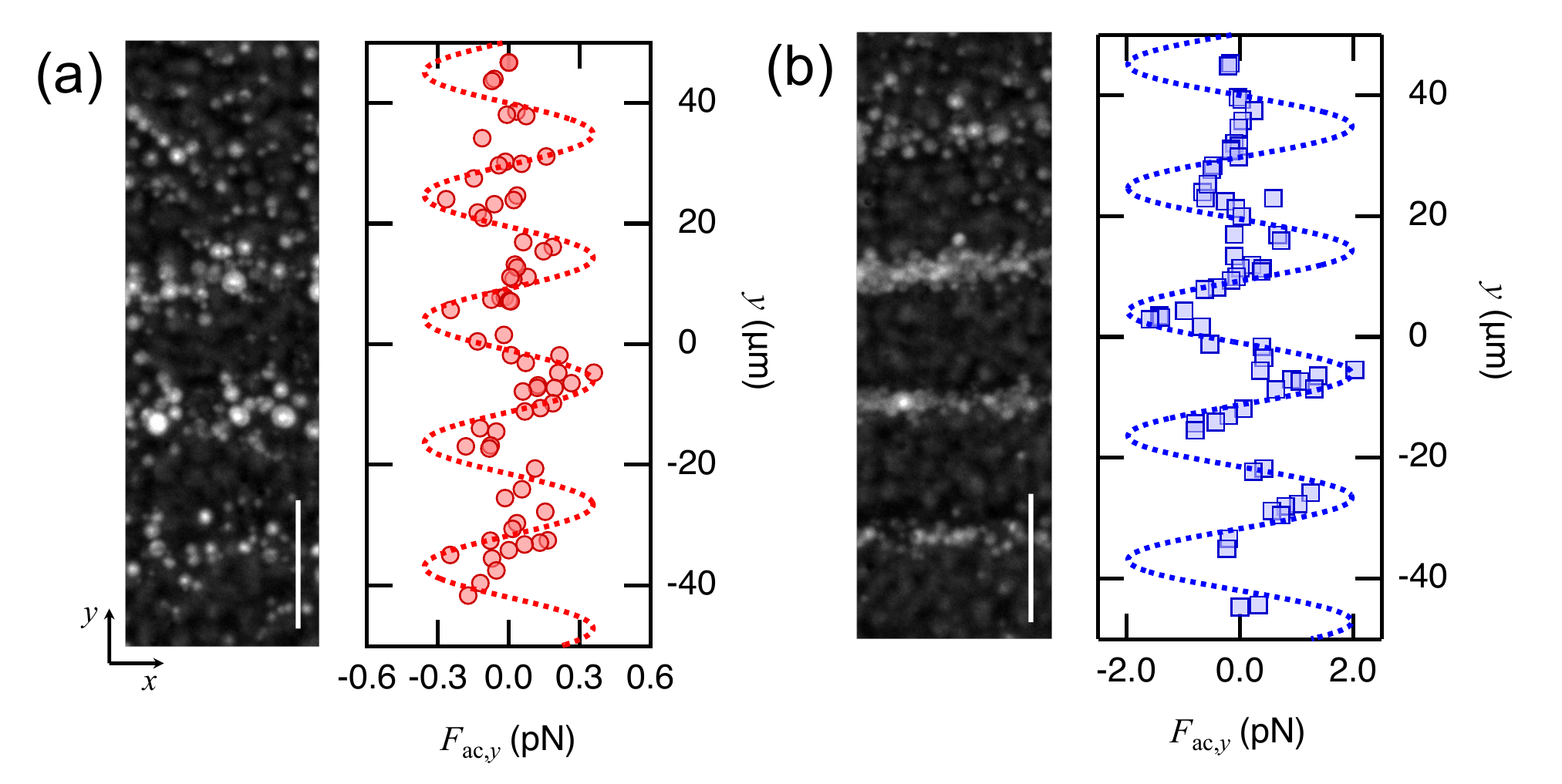}
  \caption{Effect of the superposition of SAW injected from both sides of the MF channel in a type-1 AT device. (a) single sided IDT excitation (from the top). (b) Double sided IDT excitation. The tighter aggregation of the tracer particles in (b) is apparent. The estimated maximum acoustic force $F_{\mathrm{ac}}^{\mathrm{(max)}}$ was equal to 0.4 pN for (a) and 2.0 pN for (b), respectively. The curves in the right pannels of (a) and (b) are the result of sinusoidal fitting with Eq. (2). Scale bar = 20 $\mathrm{\mu}$m.}
  \label{fig:4}
\end{figure}

Synchronously exciting the two IDTs on both sides of the MF chip is another possible way to further increase the acoustic energy density inside the MF channel. This is because the constructive interference of two incident acoustic waves will lead to a fourfold increase in $E_{\mathrm{ac}}$ in the MF channel (see Eq.~\eqref{eq:Frad} and \eqref{eq:Eac}). We tested this  using a type-1 AT device (without acoustic horn). In the experiment, we applied five 100 ms ultrasound pulses with a pulse period of 1 s with the voltage of 0.68 and 0.75 $\mathrm{V}_{\mathrm{rms}}$ to the IDTs on the left and right side (top and bottom side in Fig. 4), respectively. As shown in Fig.~\ref{fig:4}, we observed that the maximum acoustic force $F_{\mathrm{ac},y}^{\mathrm{(max)}}$ was a factor of 5 higher when two IDTs were driven (Fig.4(b)) in comparison to the case when single IDT were excited (Fig.4(a)). The tighter aggregation of tracer particles in Fig. 4(b) is apparent (see Supplementary Video 3). The observed close to fourfold enhancement of $F_{ac}^{(max)}$ in the both-side excitation case is as expected from the interference of the two acoustic waves. 

The deviation ($\sim$20\% higher) of the value from the factor of 4 is ascribed to the $\sim$10\% of difference in two RF voltages. We also suspect that the acoustic couplings were not identical on both sides. We found that the latter was more significant when the experiment was repeated with a type-2 AT device with the acoustic horns. In this case, the observed enhancement less than a factor of 2, and the excited standing waves in the MF channel excited by two IDTs were slightly shifted spatially. Nevertheless, the comparison between Fig.~\ref{fig:2} and Fig.~\ref{fig:4} demonstrates that single-sided excitation with the acoustic horn achieved the acoustic radiation force as high as double-sided excitation in the absence of acoustic horn, highlighting its effectiveness.

\subsection*{Trapping of motile cells}

\begin{figure*}[h]
\centering
  \includegraphics[width=0.9\linewidth]{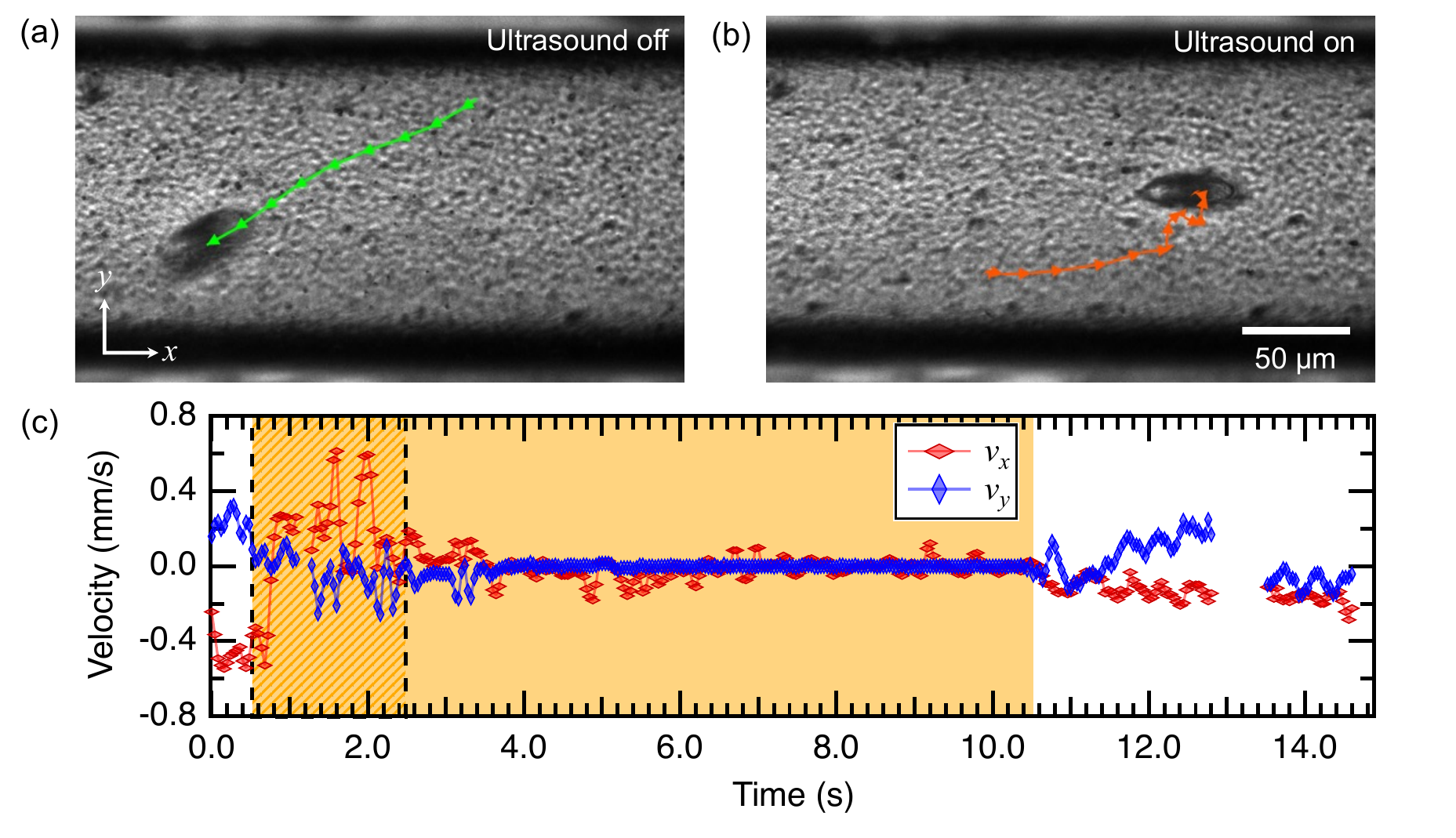}
  \caption{(a) Snapshot of the free motion of a single \textit{Tetrahymena} (the dark ellipsoid) when the ultrasound is off. The green arrows illustrate the frame-by-frame movement of the cell during $t$ = 0-0.36 s under white LED illumination. (b) Snapshot of the same \textit{Tetrahymena} when ultrasound was on. The orange arrows show the frame-by-frame movement during $t$ = 1.68-2.48 s, revealing the constriction of the cell’s motion in the vertical ($y$) direction along the pressure nodes. (c) RMS velocities $v_x$ and $v_y$, respectively in the $x$- (horizontal) and $y$- (vertical) directions of the cell's centroid in Supplementary Video 4. The ultrasound was on during $t$ = 0.5 and 10.5 s (orange shaded). The period corresponding to the 3rd column of Tab.~\ref{tab:1} is encompassed by black dashed lines and patterned in oblique orange lines. Some points are missing when the cell was along the channel wall and hidden.}
  \label{fig:5}
\end{figure*}

We next tested our AT device for trapping a moving sample, \textit{Tetrahymena}, a highly motile unicellular eukaryote. As shown in Fig.~\ref{fig:5}, it has an ellipsoidal shape with a minor radius $r_1$ $\sim$ 10 $\mathrm{\mu}$m smaller than the half wavelength $\lambda/2$ = 15 $\mu$m of 50 MHz ultrasound in water, but its major radius $r_2$ $\sim$25 $\mu$m is larger than $\lambda/2$. It moves mostly straight along its major axis.  We recorded the motion of \textit{Tetrahymena} cells for 15 s at 25 frames/s, during which we turned on the ultrasound when $t$ = 0.5 to 10.5 s by driving one IDT on the SAW chip at $V_{\mathrm{RF}}$ = 8.79 $V_{\mathrm{rms}}$ and 49.2 MHz (Supplementary Video 4). As can be observed in Fig.~\ref{fig:5} (a)-(b) and more clearly in Supplementary Video 4, there were 3 $\mu$m tracer particles dispensed in the MF channel to visualize the acoustic pressure nodes.  

\begin{table*}[h]
\small
  \caption{\ Root-mean-square (RMS) velocity when ultrasound is turn off and on, respectively.}
  \label{tab:1}
  \begin{tabular*}{\linewidth}{@{\extracolsep{\fill}}ccccc}
    \hline
     \shortstack{RMS velocity\\(mm/s)} & \shortstack{Off\\ (0 -- 0.36 s)} & \shortstack{On\\ (0.52 -- 2.48 s)} & \shortstack{On\\ (0.52 -- 10.52 s)} & \shortstack{Off\\(10.54 -- 14.6 s)} \\
    \hline
     $v_{\mathrm{tot}}^{(\mathrm{rms})}$ & 0.54 & 0.31 &  5.9$\times$10$^{-3}$ & 7.3$\times$10$^{-3}$ \\
     $v_x^{(\mathrm{rms})}$ & 0.46 & 0.29 & 5.5$\times$10$^{-3}$ & 5.7$\times$10$^{-3}$ \\
     $v_y^{(\mathrm{rms})}$ & 0.24 & 0.10 & 2.1$\times$10$^{-3}$ &  4.5$\times$10$^{-3}$\\
    \hline
  \end{tabular*}
\end{table*}

During the first 0.5 s without ultrasound, the cell on the right-hand side in the video moved obliquely at an rms velocity $v_{\mathrm{tot}}^{\mathrm{(rms)}}$ = 0.54 mm/s (see the snapshot in Fig. 5~(a) and Table 1). We first calculaed the rms velocities under ultrasound for the period between  $t$ = 0.52 and 2.48 s. When the ultrasound was turned on at $t$ = 0.52 s, its rms velocity in $y$-direction, $v_y^{\mathrm{(rms)}}$,  rapidly plunged from 0.24 to 0.10 mm/s, while in $x$-direction $v_x^{\mathrm{(rms)}}$ decreased by only 37\%, from 0.46 to 0.29 mm/s. The reduced $v_y$ is a result of the cell being trapped in the pressure nodes. Although the cell remained trapped, it could still move along the pressure nodes, which explains the lesser reduction in $v_x$. We observed that the cell hopped from one pressure node to another in the $y$-direction, resulting in a zigzag trajectory during this time. Since the cell stayed at the same focal distance as the tracer particles throughout the recording, we conclude that the cell was trapped in pressure nodes at $\sim$40 $\mu$m above the channel bottom as well. 

Beyond $t$ = 2.5 s, the motility of the cell decreased: $v_y$ is almost zero, albeit with $v_x$ fluctuating within 0.1 m/s. When the ultrasound was turned off at $t$ = 10.5 s, the cell regained its motility, especially in $y$-direction, but its speed was somewhat lower than that before applying ultrasound. The same recording also captured the motion of a smaller cell on the left-hand side (minor and major radii equal to 8.6 and 22 $\mu$m, respectively). During $t$ = 3.5 -- 10.5 s, despite the low motility of the larger cell, the smaller cell remained somewhat agile and moved away from the recording area at $t$ = 8.5 s while the ultrasound was still on, probably because of the smaller acoustic stress with a smaller $r_1$.

\begin{figure*}[h]
\centering
  \includegraphics[width=0.9\linewidth]{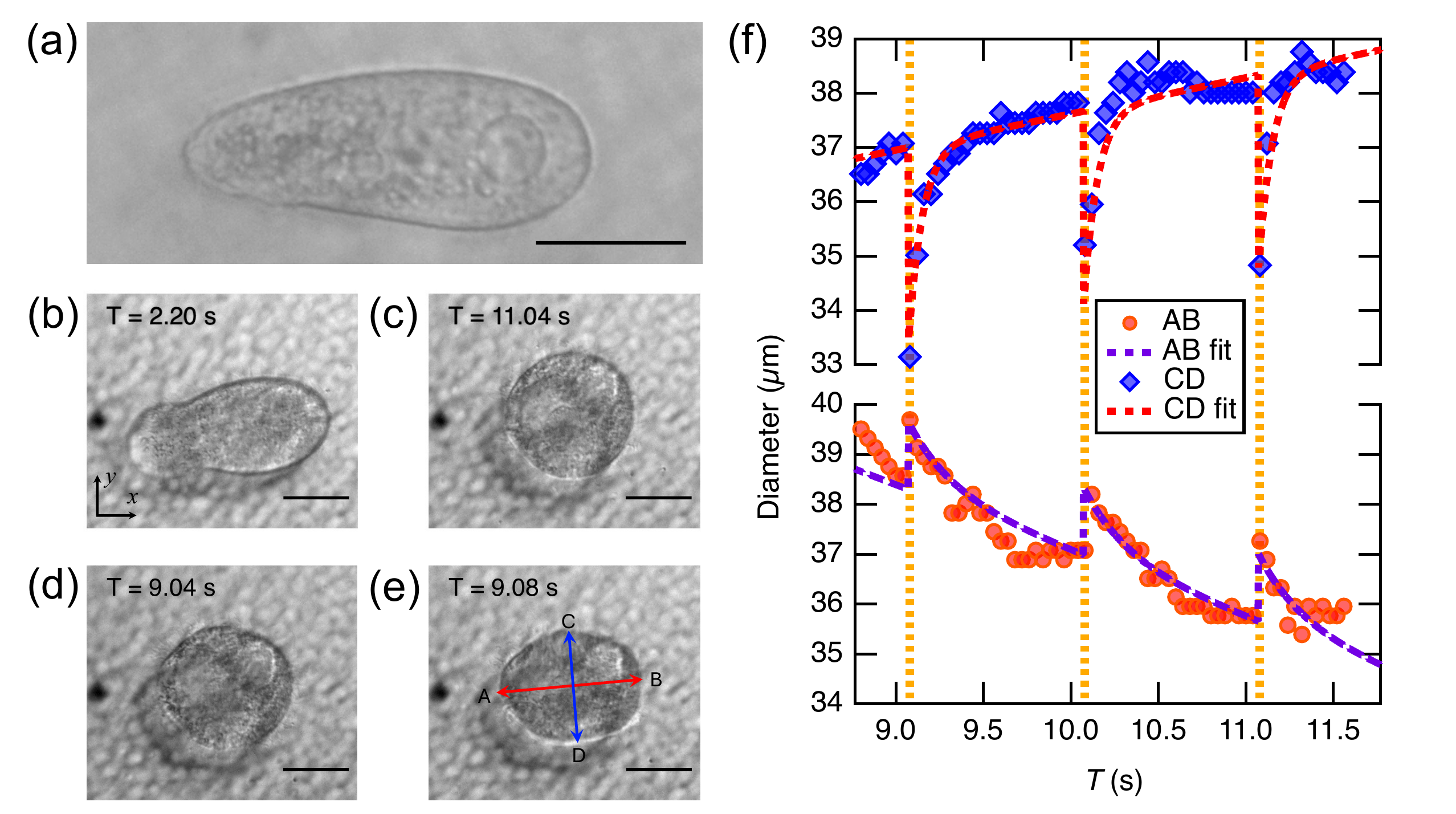}
  \caption{Time evolution of the \textit{Tetrahymena} under UV illumination and acoustic pulses. $T$ is the time measured from $T_0$ = $\sim$12 min after starting the UV illumination. Acoustic pulses were applied from $T_0$ with an approximate interval of 1 s. (a) Intact \textit{Tetrahymena} after $\sim$5 min of UV illumination.  Between 2.20 and 11.04 s after $T_0$, respectively (b) and (c), the cell morphology changed from an ellipsoid into a sphere. Note that (b-e) are of the same cell but (a) is different. from 9.04 to 9.08 s after $T_0$, respectively (d) and (e), the deformation of the cell was observed. (f) Vertical and horizontal diameters of the \textit{Tetrahymena} between 8.5 and 12 s after $T_0$. During this duration, a large deformation of the cell was observed but not before or after (see Supplementary Information). Scale bars are 20 $\mu$m.}
  \label{fig:6}
\end{figure*}

The order-of-magnitude estimate of the propulsion $F_\mathrm{p}$ of the cell during the free motion ($t$ = 0 - 0.5 s) is equal to 0.1 nN by applying Eq. (4) with $v_{\mathrm{tot}}$ = 0.54 mm/s and using $r_1$ as the radius. In comparison, the maximum acoustic force $F_{\mathrm{ac}}^{\mathrm{(max)}}$ on the cell with $r_1$ of 10 $\mu$m is of the order of tens of nN, according to the scaling of the force observed for PE tracers (Since $r_2$ > $r_1$, the actual force is somewhat larger), and much larger than $F_\mathrm{p}$. Therefore, the rapid reduction of $v_y$ upon the application of ultrasound at $t$ = 0.5 s and the vertical confinement of the cell motion along the acoustic pressure node appear to be plausible. The restrained motion observed in the horizontal direction could be associated with the application of acoustic stress, although we were not able to resolve any compression of the cell in the present experiment due to the rapid motion and the limited optical resolution (pixel size = 0.4 $\mathrm{\mu}$m).

\subsection*{Deformation of inactivated cells}

When \textit{Tetrahymena} was illuminated by UV instead of white LED light, they gradually lost their motility and eventually died.\cite{Fuma:2003:1} 
After 
$T_0$ $\equiv$ 12 min of UV illumination, the cell began to change its shape from an ellipsoid (Fig. 6(a) and (b)) to a sphere (Fig. 6(c)). Note that in the figure $T$ = 0 s corresponds to $T_0$. From $T_0$ ($T$ = 0 s), 3 ms long ultrasound pulses were applied with a period of 1 s and an amplitude of 45.1 V$_{\mathrm{rms}}$ at 49.2 MHz.  As displayed in Fig. \ref{fig:6} (d), (e), and (f) (also see Supplementary Information), strong deformation induced by acoustic pulses was observed between $T$ = 8.5 and 12 s: compression in the vertical direction and elongation in the horizontal direction. Fig. \ref{fig:6} (f) shows that the  deformation was followed by a slow recovery. This is a typical behavior of a viscoelastic material. A detailed discussion can be found in Section C in Supplementary Information. The observation of such large deformation only after a certain amount of time suggests that the stiffness of the cell gradually decreased during its morphological change. Before $T$ = 8 s, it is likely that the cytoskeleton of \textit{Tetrahymena} sustained its integrity against the much lower acoustic radiation pressure for trapping (estimated to be $\sim$250 Pa in this experiment), with negligible deformation in comparison with the optical resolution ($\sim$0.5 $\mu$m) in the experimental setup. The variation in the cellular membrane stiffness during the morphological alternation of \textit{Tetrahymena} has not yet been fully investigated and is beyond the scope of this study.

\begin{figure}
    \centering
    \includegraphics[width=\linewidth]{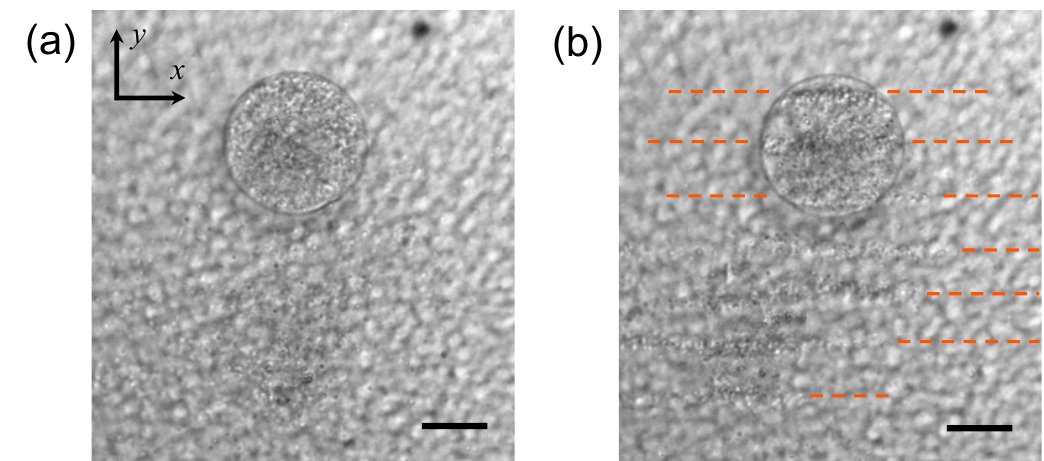}
    \caption{(a) A dead Tetrahymena cell that morphed into a sphere after $\sim$12 min of UV irradiation. (b) Aggregation of intracellular organelles in the dead Tetrahymena in (a), caused by the acoustic radiation force towards the pressure nodes (orange dashed lines). Scale bars are 20 $\mu$m. Scale bars are 20 $\mu$m.}
    \label{fig:7}
\end{figure}

The lack of large deformation after $T$ = 12 s (see Supplementary Information) indicates that the reduced density inside the cell due to the purported emission of its intracellular components, partly aided by the acoustic pulse induced deformation (See Supplementary Video 5). Interestingly, at this point, we observed an aggregation of remaining substances inside the cell along the pressure nodes (dark horizontal lines marked by orange dashed lines in Fig.~\ref{fig:7}(b)) in response to acoustic pulses. Instantaneous movement of these substances and the build-up of the aggregation by the acoustic pulses are visible in Supplementary Video 6. In fact, the period is approximately equal to $\lambda /2$ of the ultrasound as expected (See Figure 2). This showcases that the acoustic wave penetrated into the cell and affected the intracellular substances.

We found that fine particles suspended in the fluid are visible in Fig.~\ref{fig:7} (a). They were believed to be the intracellular substances from the cells that had previously decomposed after membrane rupture (one such event was recorded in Supplementary Video 6 from $T$ = 20.60 s). The bottom-left part of Fig.~\ref{fig:7}(b) shows the ultrasound-induced aggregation of these particles in the fluid outside the cell as well.

\subsection*{Deformation of agarose hydrogel beads}

\begin{figure}[h]
\centering
  \includegraphics[width=\linewidth]{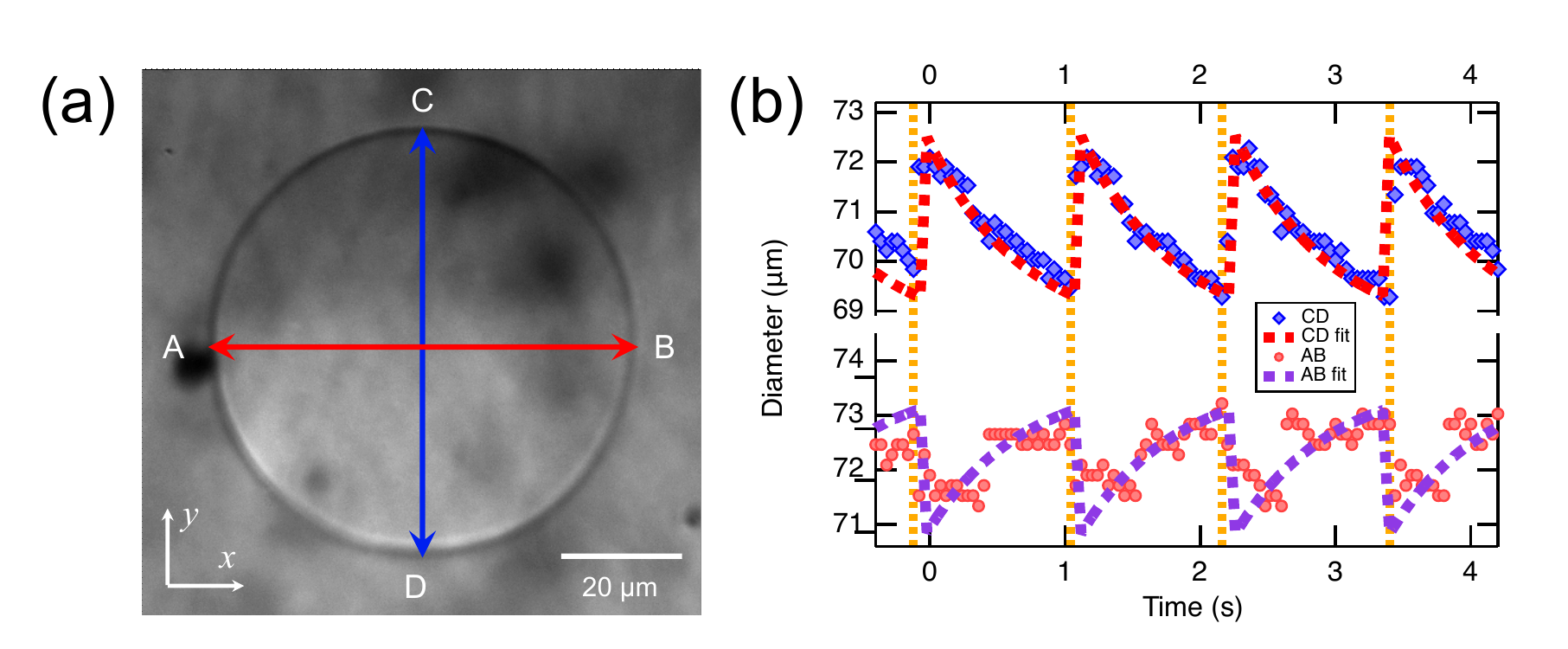}
  \caption{(a) A snapshot of a hydrogel bead in the MF channel. (b) Variation of the hydrogel bead diameters in the horizontal (red circles) and vertical (blue diamonds) direction when the ultrasound pulses with 50 ms pulse duration was applied (orange vertical lines). The chain curves are the fits with a time constant of 0.7 s.}
  \label{fig:8}
\end{figure}

Finally, we discuss the experiment for testing the deformation of the viscoelastic material using hydrogel beads as the one in Fig.~\ref{fig:8} (a). We used a device having a 500 $\mathrm{\mu}$m wide MF channel to accommodate these large beads. After a small number of beads were suspended in DI water and infused into the MF channel,  ultrasound pulses were applied to the device, which were produced by $V_{\mathrm{RF}}$ = 15.8 $V_{\mathrm{rms}}$ with a pulse duration $\Delta{t}$ of 50 ms and a period of 1.1 s. 

The observed deformation of the beads, along CD in the vertical direction (in the $y$ direction) equal to +3\% and along AB in the horizontal direction (in the $x$ direction) equal to -2\%, are shown in Fig.~\ref{fig:8} (b). The instantaneous deformation upon applying acoustic pulses (indicated by orange vertical lines) and its slow recovery show the viscoelastic response of hydrogel.\cite{Roberts:2011} The recovery time $T$ of the hydrogel was found to be $\sim$0.7 s (red and purple dashed curves in Fig. 7 show the exponential fit).

The compression of the bead was along the MF channel and perpendicular to the $F_x^{rad}$ on tracers [Fig.~\ref{fig:2})]. We consider this to be an indication that the acoustic standing wave was disturbed by the bead which was much larger than $\lambda$. In fact, we also observed an event when the direction of the acoustic force was oblique when multiple beads were close to each other. The full understanding of the observation would require three-dimensional analysis of the steady-state acoustic wave distribution. However, we found that the result of a finite element simulation (See Supplementary Fig. S5) conducted with a simplified, 2-dimensional model, is qualitatively in line with the experiment: there is a case when the pressure maxima on the bead are on the horizontal side.  

Noting that $\Delta T$ is much shorter than $T$ and the size of the hydrogel beads are two times larger than the acoustic wavelength, we evaluate the order of magnitude of the bead's elastic modulus $E$ by the relationship, $(\Delta T/T) P_{\mathrm{rad}}/|\epsilon|$. Here, $|\epsilon|$ = 0.02-0.03 is the strain, given by the ratio of the deformation to the diameter of the hydrogel bead (see Section D in Supplementary Information), and  $P_{\mathrm{rad}}$ is equal to $\sim$100 Pa. We found $E$ in the range of 250-350 Pa. This estimated value appears to be consistent with some reports.\cite{Roberts:2011} Nevertheless, values\cite{Yan:2009} that are orders of magnitude higher were also reported for hydrogel with a similar concentration of agarose. Such variation in the material parameters is likely in part ascribed to water-induced swelling of hydrogel \cite{Roberts:2011} and requires further study including measurements using other methods or ultrasound of a longer wavelength that matches better the present sample.

\section*{Conclusions}
In summary, we show that a high acoustic pressure amplitude up to a few MPa can be achieved in a hybrid acoustic tweezing device comprising a SAW chip and a Si MF chip. This was a result of realizing an acoustically tight bonding between the two chips and adding acoustic horns to the Si MF chips. The fabricated device enables the trapping of highly motile unicellular eukaryotes and the viscoelastic testing of hydrogel samples. Acoustic coupling efficiency is expected to be further enhanced by optimizing the acoustic coupling across the chip-to-chip interface. This can be achieved by refining the acoustic horn, taking into account the three-dimensional geometry of the device. The focus should shift from bulk-acoustic excitation to surface acoustic waves.

Compared to monolithic AT devices, our separate device construction offers significant advantages. It allows for shaping the Si chip using DRIE to enhance acoustic coupling and enables the disposal of Si MF chips to prevent contamination while reusing the more expensive SAW chips. With the present device design and construction, the reuse of SAW chips is still limited to 3-5 times  due to the wear of IDT electrodes while removing resist and cleaning chips. Nevertheless, we consider it straightforward to improve this by material choice and/or protective coating over the electrodes.\cite{Freudenberg:2001} 
With such a hybrid construction, our AT will be advantageous to practical applications that demand the analysis of a large variety of samples, where cross-contamination must be avoided. 

Further improvement of the device for such experiments will be to mount the assembled AT devices into a holder with a standardized MF interface holders,\cite{vanderLinden:2020} some of which seem to be commercially available. In comparison with the present manually sealed experiments, such arrangement will improve the throughput of the experiments. 

The observed diminished motility of the cell after trapping and the observed formation of acoustic standing waves inside the inactivated \textit{Tetrahymena} cells suggest that the acoustic waves do affect the cell as well as its internals. Although we have not been able to observe the mechanical disturbance to the cell by the acoustic wave, we suppose that increasing the acoustic radiation force through further optimization of the device structure and assembly method will introduce a novel modality to mechanobiological experiments in addition to the widely used optical tweezers \cite{Guck:2001, Cojoc:2004, Lee:2023} in a non-contact way and in contrast to the experiments conducted by AFM (atomic force microscopy) or by using the hydrodynamic force.\cite{Hao:2020}

\section*{Author contributions}
S.T. conceived the experiment. S.J. fabricated the device, optimized the device assembly methods, conducted the AT experiments, and analyzed the data in close communication with S.T. S.T. conducted the FEM simulation. S.J. and S.T. wrote the manuscript.

\section*{Conflicts of interest}
There are no conflicts to declare.

\section*{Acknowledgements}

The authors thank Yagyik Goswami, Earni Varaha Trinadharao Sornapudi, and G. V. Shivashankar for discussions on viscoelasticity and potential applications of AT, and Roderick Y. H. Lim and colleagues in the Lim Lab for discussions on the AT applications in nanobiology. The authors acknowledge Konstantins Jefimovs, Dario Marty, Zhitian Shi, Konrad Vogelsang, Anja Weber, and Christopher Wild for the support and inputs on the device  fabrication in the Park Innovaare (PiA) Cleanroom for Optics and innovation (PICO). The project was partially supported by the Swiss Nanoscience Institute Ph.D. School (Project No. P2007).



\balance


\bibliography{references} 
\bibliographystyle{rsc} 



\section*{Supplementary material}

<div align="center">

# Supplementary Information
## Surface-acoustic-wave driven silicon microfluidic chips
## for acoustic tweezing of motile cells and viscoelastic microbeads

</div>

<div align="center">

Shichao Jia[*1,2,3] and Soichiro Tsujino[‡1,2]

[1]*PSI Center for Life Sciences, Paul Scherrer Institut,*
*Forschungsstrasse 111, Villigen-PSI, 5232, Switzerland.*
[2]*Swiss Nanoscience Institute, University of Basel, Klingelbergstrasse 8, Basel, 4056, Switzerland.*
[3]*Biozentrum, University of Basel, 4056, Spitalstrasse 41, Switzerland.*

</div>

Electronic mail: ∗ shichao.jia@psi.ch; ‡ soichiro.tsujino@psi.ch

**A. Spatial concentration of the acoustic radiation by acoustic horns in the type-2 AT device**

**B. Deformation of a morphologically changing *Tetrahymena* cell**

**C. Deformation of viscoelastic materials**

**D. Two-dimensional finite element simulation of the acoustic wave in the MF channel with a large bead**

**Supplementary Video 1: Acoustically tweezing of PE particles in type-1 AT device (w/o horn)**

**Supplementary Video 2: Acoustically tweezing of PE particles in type-2 AT device (w/ horn)**

**Supplementary Video 3: Acoustically tweezing of PE particles in type-1 AT device (w/o horn) by synchronously driving two IDTs**

**Supplementary Video 4: Freely swimming of a *Tetrahymena* when ultrasound is off and its constrained motion when ultrasound is on**

**Supplementary Video 5: Aggregation of intracellular substances by acoustic wave**

**Supplementary Video 6: Acoustic deformation of a *Tetrahymena* during its morphological alternation**

**Supplementary Video 7: Horizontal entrapment of a *Tetrahymena* in type-2 AT device**

## A. SPATIAL CONCENTRATION OF THE ACOUSTIC RADIATION BY ACOUSTIC HORNS IN THE TYPE-2 AT DEVICE

### 1. Concentrated acoustic radiation in the MF channel

As shown in Fig. 1 in the main text, the tracer particles visualize the acoustic nodal lines after applying acoustic pulses that are produced on the SAW chip and coupled to the Si MF chip. In the type-2 AT device, the acoustic horn concentrates the ultrasound within 200 $\mu$m between the two vertical dashed lines in the Supplementary Figure S1. This was recorded after applying five consecutive 100 ms pulses at 0.68 $V_{rms}$ and 49.2 MHz to the device, wherein the PE tracers distributed randomly at the beginning. The acoustic wave is concentrated in this region, where PE tracers are aggregated along the horizontal nodal lines with a spacing of $\sim$16 $\mu$m, but the PE distribution is random outside this region. The acoustic nodes are again visible in the region 100-200 $\mu$m away from the central part, indicating a side-lobe distribution, where the acoustic radiation pressure is much weaker than the central part.

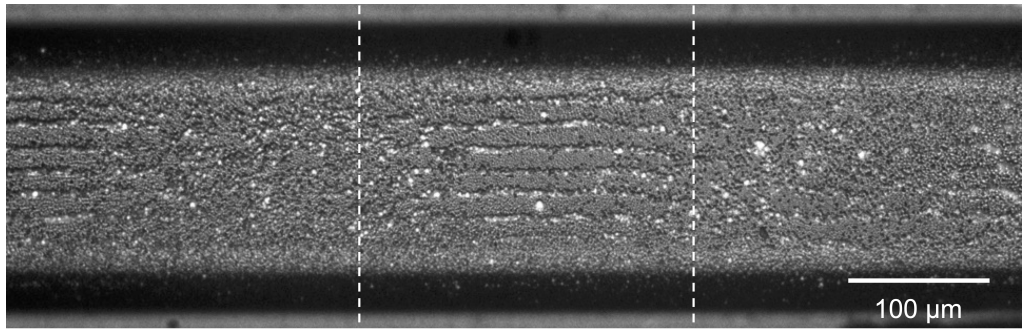

FIG. S1: Aggregation of PE tracer particles in the type-2 AT device with the acoustic horn.

### 2. Horizontal entrapment of motile cells in type-2 AT device

We observed that in addition to the vertical direction, the motion of the motile cells along the horizontal direction was also constricted within the $\sim$200 $\mu$m long region in the MF channel where the acoustic radiation is concentrated. As shown by the orange arrows in Fig. S2, while the ultrasound was on, in addition to the vertical hopping of the cell between acoustic pressure nodes, it also turned back once it reached the boundary marked out by the vertical dashed lines while moving along the acoustic nodes in the horizontal direction. When the ultrasound was turned off, the cell again started swimming freely (green arrows).

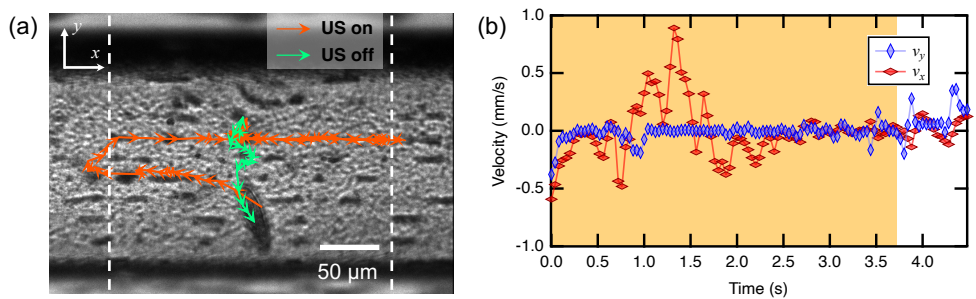

FIG. S2: (a) Snapshot of a *Tetrahymena* cell and its precedent trajectory when ultrasound (US) was turned on at 8.13 Vrms and 49.2 MHz (orange arrows) and off (green arrows). Vertical dashed lines indicate the edges of the region projecting from the tip of the acoustic horn, where the acoustic radiation is concentrated. (b) Frame-by-frame velocities of the cell analyzed from the Supplementary Video 7 as shown in (a). The orange shade represents the period when the ultrasound was turned on.

This is a consequence of the acoustic radiation force at the boundaries where it creates horizontal entrapment since the acoustic radiation force is proportional to the spatial derivative of the acoustic energy density[1].

## B. DEFORMATION OF A MORPHOLOGICALLY CHANGING *TETRAHYMENA* CELL

After the cell was illuminated by UV light for over 12 min, it morphed from a elipsoild-like shape into a sphere at ~12 min 10 s (Fig. 6 (c) in the main text). Between these two time points, we observed a strong deformation of the cell upon the application of the ultrasound pulses, as shown in Supplementary Video 5.

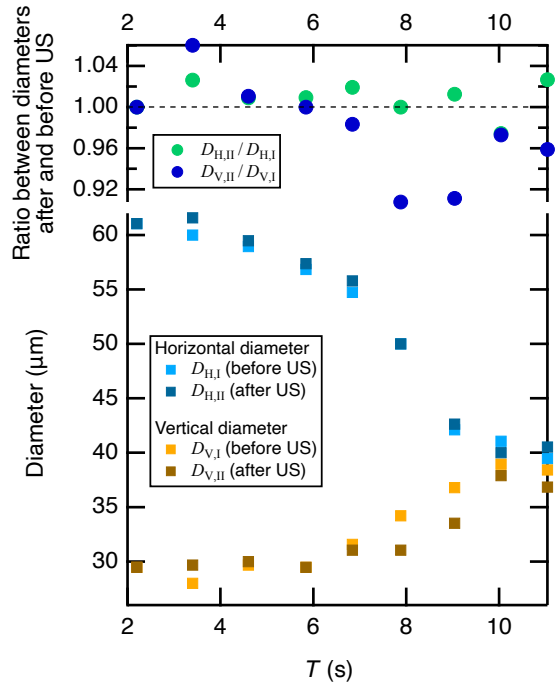

FIG. S3: Time evolution of the cell diameters between 8.8-11.7 s and of the ratio between the diameters after and before applying ultrasound (US)

In the subsection "Deformation of the inactivated cells" of the main text, we discussed the presumably UV-induced morphological change of *Tetrahymena* and the deformation of a *Tetrahymena* during this stage.

In Fig. S3, we show the overall trends of the decreasing horizontal diameter, $D_{H,i}$, and the increasing vertical diameter, $D_{V,i}$, where $i =$ I, II denotes just before and just after applying US pulse, respectively. At $= 7.88$ and $9.04$ s, $D_{V,II}/D_{V,I}$ was found to be equal to 0.92. As decribed in the main text, this proves significant vertical compression despite the cell's own vertical elongation, suggesting its reduced stiffness. The observation of such large deformation merely at these two events implies the varying stiffness of the cell during its morphological change.

Fig. 6 in the main text and Supplementary Video 6 record such deformation, as in, the cell was vertically (along CD) compressed and horizontally (along AB) elongated. The duration of the applied acoustic pulses was $\Delta T = 2$ ms, and the excitation voltage was 32.5 $V_{rms}$. The acoustic radiation pressure $P_{rad}$ is estimated to be ~600 Pa by assuming that the acoustic contrast factor for *Tetrahymena* in water could be approximated by that of PE. CD is not exactly perpendicular to the pressure nodal lines as previously visualized by PE tracer particles. This suggests a slightly different acoustic standing wave distribution in the present experiment because the cells are much larger than PE tracer particles. The observed slow recovery from deformation, as plotted in Fig. 6(f) in the main text, is a signature of the viscoelasticity of the cell. The fitting shows the exponential recovery with its time constant $T_c$ equal to ~0.1 s (0.12 s and 0.08 s in the horizontal and vertical direction). From the estimated $P_{rad}$, the ratio $\Delta T/T_c$, and the observed deformation (compression of 11 %), the elastic modulus of this cell is estimated to be ~140 Pa (see also Section D) and orders of magnitude smaller than the value of an intact eukaryotic cell of the order of 10-100 kPa.[2] In literature, exposing *Tetrahymena* to cytotoxic substances was found to result in the reorganization of the actin filaments in the cell,[3] which maintain its structural integrity.[4] Therefore, we consider that the UV irradiation may have also caused similar disintegration of the cell. In fact, excess UV irradiation has been shown to be cytotoxic to *Tetrahymena* as well.[5]

## C. DEFORMATION OF VISCOELASTIC MATERIALS

It is expected for the viscoelastic bead to be compressed in the direction of the acoustic force and elongated in the perpendicular direction. For the quantitative analysis of the experiment, a three-dimensional analysis of the acoustic wave distribution and the deformation of the hydrogel bead is required, which is beyond the scope of the present work. For the order-of-magnitude analysis of the acoustic force and elastic modulus $E$ of hydrogel, we consider the one-dimensional deformation of hydrogel: the strain $\epsilon$ of the deformation is given by the solution to the following equation (Voigt model, see Ref. [6]),

$$\frac{d\epsilon}{dt} + \frac{\epsilon}{T_c} = -E\epsilon + P_{rad}(t). \tag{S1}$$

where $T_c$ is the relaxation time, $E$ is the elastic modulus, and $P_{rad}(t)$ is the time dependent acoustic radiation pressure. For the hydrogel beads that are much larger than the acoustic wavelength, the acoustic force cannot be described by Eq. (2) in the main text due to its small-size limit, as is well-known in literature. [7–11] For quantitative assessment, three-dimensional analysis is required. However, we approximated the acoustic radiation pressure equal to ∼100 Pa from $V_{RF}$ and assumed the acoustic contrast factor of the hydrogel equal to that of PE. The solution of the above equation gives the fitting curves in Fig. 7, and the peak deformation is equal to

$$\epsilon(t = \Delta t) \simeq \frac{\Delta T}{T_c}\frac{P_{rad}}{E}. \tag{S2}$$

## D. TWO-DIMENSIONAL FINITE ELEMENT SIMULATION OF THE ACOUSTIC WAVE IN THE MF CHANNEL WITH A LARGE BEAD

In the experiment using a 70-$\mu$m-diameter hydrogel bead, the observation of the compression in the $x$-direction (along the channel) indicates that the acoustic standing wave was formed in an peculiar way. This is likely due to the fact that the sample size was more than 4 times the acoustic half wavelength, $\lambda/2 \sim 15$ $\mu$m; the scattering by the sample itself has a large influence on the standing wave pattern. As a reference, we simulated the acoustic wave distribution in our microfluidic channel by a finite element simulation software (COMSOL). We modeled the sample in a two-dimensional geometry, with the acoustic wave uniformly incident from the left-edge of the left-side part of the Si block. The width of the fluidic channel equal to 500 $\mu$m was nominally the same as the sample. However, the width of the Si part was assumed to be only 100 $\mu$m. The length of the model was taken until the further increase does not influence the acoustic wave distribution significantly. Standard values were assumed for the sound speeds and the mass densities of silicon and water. We modeled the bead with the sound speed and the density slightly higher than water ($c^{(bead)} = 1.55\times10^3$ m/s and $\rho^{(bead)} = 1.05\times10^3$ kg/m$^3$, respectively). The soft-boundary condition was assumed at the right-side edge of the right-side silicon block. The top and bottom edges of the geometry were assumed to satisfy the periodic boundary condition for respective part.

The acoustic standing wave in the water channel for the acoustic frequency around 50 MHz is formed when $\lambda$ is equal to 32 or 33 $\mu$m with the pressure peak at the interface between Si and water. The large acoustic pressure $p_{ac}$ in the MF channel is attained when the acoustic wave is in resonance within the silicon block as well. For the assumed two-dimensional geometry of the bead (modeled as a disc at the center of the fluidic channel), one of the resonances near 50 MHz occurs at 48.285 MHz, which is in agreement with the experiment within a few percents. The distribution of the absolute acoustic pressure amplitude (arb. units) in this condition is shown in Fig. S4 (b). As shown in the expanded view near the bead in Fig. S4 (c), the acoustic wave is weakened in the region where the bead shadows the incident wave, and the result is a large $p_{ac}$ on the top and bottom of the bead. This is qualitatively in agreement with the experiment. For comparison, when the acoustic parameters of the bead are equated to those of water, the resonance condition at 48.2675 MHz results in the standing wave as shown in Fig. S4 (e)-(f), which is uniform along the channel (in the vertical direction in Fig. S4).

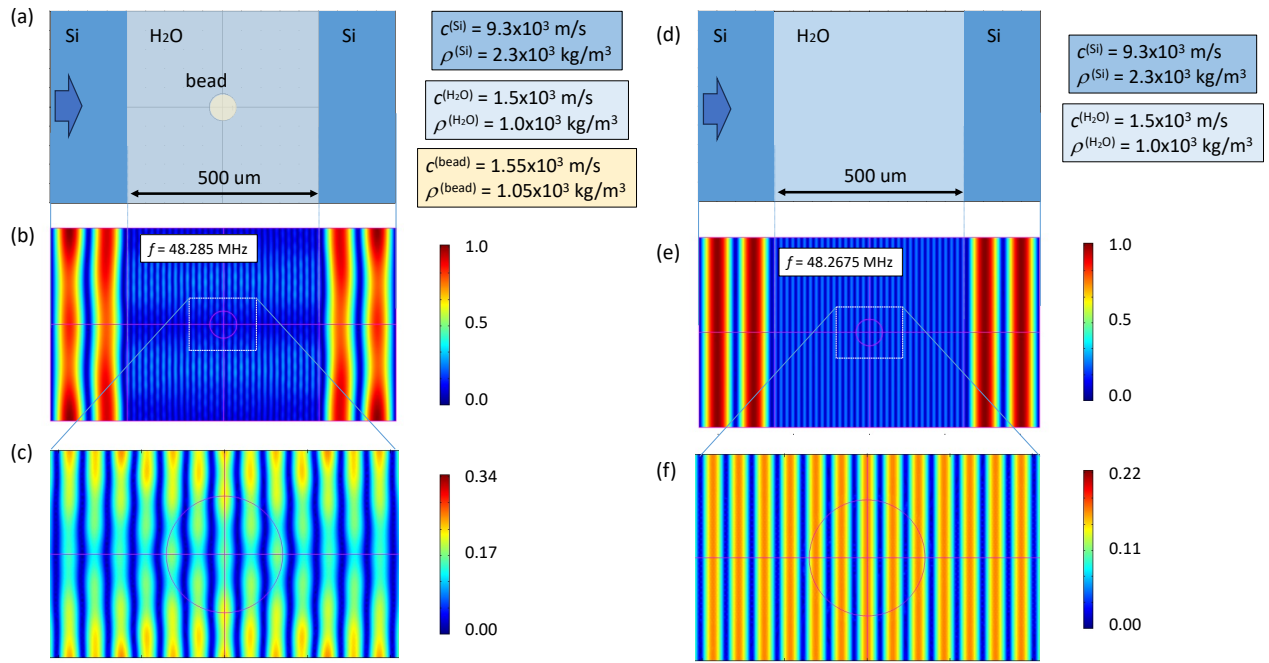

FIG. S4: Two-dimensional finite element simulation of the acoustic resonance in the silicon microfluidic channel. (a-c) show the case where a bead is placed at the center of the channel, and (d-f) show the uniform channel case. The fluidic channel with the width of 500 $\mu$m is filled with water. The arrow in (a) and (c) indicates the incident acoustic wave from the left-most edge of the left-side silicon block.



\end{document}